\documentstyle[aps,prb,preprint]{revtex}

\begin{document}
\title{Two-Carrier Transport in Epitaxially Grown MnAs}
\author{J. \ J. \ Berry, S.\ J.\ Potashnik, S.\ H.\ Chun, K.\ C.\ Ku, P.\ Schiffer, and N.\ Samarth\cite{email}}
\address{Department of Physics and Materials Research Institute, \\ The Pennsylvania State University, University Park, PA 16802}
\date{\today}
\maketitle

\begin{abstract}
Magneto-transport measurements of ferromagnetic MnAs epilayers grown by molecular beam epitaxy reveal the presence of both positive and negative charge carriers. Electrical transport at high temperatures is dominated by holes, and at low temperatures by electrons. We also observe distinct changes in the magnetoresistance associated with the transition between the electron- and hole-dominated transport regimes. These results are of direct relevance to MnAs/semiconductor hybrid heterostructures and their exploitation in electronic and optical spin injection experiments.
\end{abstract}
\pacs{75.50.-I, 72.25.Rb, 72.25.Ba}

% Use the \preprint command to place your local institutional report
% number on the title page in preprint mode.
% Multiple \preprint commands are allowed.
%\preprint{}
%\bigskip
%\vspace*{0.5in}
\newpage
Contemporary interest in hybrid ferromagnet/semiconductor heterostructures is driven by their relevance to ``spintronic'' semiconductor applications that rely on spin injection from a ferromagnet (or paramagnet) into a semiconductor.\cite{spininject} In this context, epitaxially grown layers of MnAs -- a room temperature ferromagnet - are of active interest since they can be successfully integrated through molecular beam epitaxy (MBE) with technologically relevant semiconductors such as GaAs, \cite{tanaka1} Si,\cite{tanaka2} and ZnSe.\cite{berryapl} Although the magnetic and structural properties of bulk MnAs samples were initially studied several decades ago,\cite{menyuk} a comprehensive understanding of the structural and magnetic properties of technologically relevant MnAs epilayers has begun to emerge only recently. These recent studies show that the physical properties of the epitaxial layers can differ significantly from those of bulk samples. For instance, since MnAs epilayers grown on (001) GaAs exhibit strain-stabilized co-existing phases of $\alpha$-MnAs and $\beta$-MnAs (in apparent violation of the Gibbs phase rule), \cite{kaganer} the magnetic behavior is quite complex.\cite{chunapl} More importantly, the electrical transport properties of epitaxial MnAs -- of critical importance in ``spintronics'' applications such as electrical spin injection -- are poorly understood \cite{govor} and there are no studies of the temperature or field dependence of the magneto-resistance and Hall effect in MnAs epilayers in the literature. Here, we report a detailed study of the longitudinal resistivity ($\rho _{xx}$) and the Hall resistivity ($\rho _{xy}$) as a function of temperature and magnetic field in MnAs epilayers grown on both (001) GaAs and (001) ZnSe. Our data indicate that both electrons and holes contribute to the electrical transport, and that this co-existence leads to a low temperature regime of large positive magneto-resistance (MR).

This paper focuses on a set of four MnAs epilayer samples grown by MBE: samples A, B, and C (with respective thickness 51nm, 41nm, and 46nm) are grown directly on (001) GaAs, while sample D (30 nm thickness) is grown after the deposition of a 700 nm thick ZnSe buffer layer on (001) GaAs. Details about the growth and structural properties of these samples have been reported elsewhere.\cite{chunjvst} All the samples show the presence of single crystal, hexagonal ferromagnetic MnAs with a Curie temperature T$_{C} \sim $320 K, but differences in growth conditions result in varying crystalline orientation and crystal quality. X-ray diffraction measurements show that samples A and B have the c-axis of the hexagonal unit cell in the (001) plane of the underlying substrate (``type-A'' orientation), while for samples C and D, the c-axis is at an angle with the (001) plane (``type B'' orientation).\cite{tanaka1} These structural differences lead to variations in microstructure that are reflected in the residual resistivity ratio (RRR), defined here as the ratio $\rho _{xx}$(350K)/ $\rho _{xx}$(2K). Samples A, B, C, and D show RRR = 195, 109, 75, and 5.7, respectively, which correlate systematically with observed trends in the electrical transport, as discussed below.

Each sample is patterned into a Hall bar using conventional photolithography and a wet chemical etch (K$_{2}$Cr$_{2}$O$_{7}$:H$_{2}$SO$_{4}$:H$_{2}$O). Electrical contacts are made using gold wire leads and indium solder. Temperature-dependent MR and Hall effect measurements are carried out using AC techniques in an external field of up to 14 T, applied perpendicular to the epilayer plane. The temperature- and field-dependent magnetization is measured using a commercial DC superconducting quantum interference device (SQUID) magnetometer. We note that, in all samples studied here, the easy axis of the magnetization is in the plane of the epilayer, but the electrical transport properties within the plane show negligible anisotropy.

Figure 1(a) compares the temperature-dependence of the sample resistivity ($\rho _{xx}$) in one of the high quality epilayers (sample B) at B = 0 T and B = 14 T. We note that resistivity indicates metallic behavior and that the high field MR changes from positive to negative with increasing temperature. Figure 1(b) depicts the temperature variation of the high field MR (defined here as MR=($\rho _{xx}$(B)$-\rho _{xx}$(0))/$\rho _{xx}$(0), with B = 14 T) for all four samples; the inset to this figure displays the temperature variation of the zero-field conductivity ($\sigma _{xx}$). The behavior of the high field MR broadly demarcates three regimes, providing a common framework for discussing the Hall and MR data in all the samples. These three regimes approximately span the following temperature ranges: region I (2-80K) displays a positive MR; region II (80-250K) shows a very weak, negative MR; and region III (250-360K) shows a modest negative MR with a peak near the Curie temperature (T$_{C}$). We note that the size of the positive MR at low temperatures increases with the RRR. We now focus on a more detailed examination of both the Hall resistivity ($\rho _{xy}$) and the MR of sample B (Figure 2).

We begin with a discussion of the Hall effect, first recalling that the Hall resistivity in a ferromagnet is given by:\cite{hurd}
\begin{equation}
\rho _{xy}=R_{H}B+R_{A}\mu _{0}M,  \label{1}
\end{equation}
where $R_{H}$ is the ordinary Hall coefficient, $R_{A}$ is the ``anomalous'' (or extraordinary) Hall coefficient and $M$ is the sample magnetization. Experimental studies of ferromagnets indicate that the anomalous Hall effect in many systems is well-described by the following empirical ansatz:
\begin{equation}
R_{A}\sim a\left[ \rho _{xx}\left( B,T\right) \right] ^{\gamma },  \label{2}
\end{equation}
wherein the exponent $\gamma $ is determined by the dominant scattering mechanism, with values ranging from 1 to 2 for the ``skew-scattering'' and ``side-jump'' mechanisms, respectively.\cite{berger} The constant $a$ is usually considered to be independent of magnetic field and temperature. We now examine the Hall effect in sample B within this phenomenological context.

At the lowest temperatures in region I (Figs. 2(a), (b)), $\rho _{xy}$ shows a clearly non-monotonic magnetic field dependence that has a positive slope at low fields and a negative slope at high fields. We note that in region I, the sample resistivity $\rho _{xx}$ is relatively small (Fig. 1(a)). Furthermore, as shown in Fig. 3 (a), the magnetization saturates beyond B $\sim $ 2 T. According to Eqs. (1) and (2), the magnetic field variation of $\rho_{xy}$ for B $>$ 2T is then dominated by the ordinary Hall effect, enabling a credible determination of both the sign and concentration of the majority carriers from high field Hall data. Hence, the negative slope of $\rho _{xy}$(B) at high field indicates electron-dominated transport at the lowest temperatures. As the temperature increases, however, the high field Hall coefficient switches sign, indicating a transition from electron- to hole-dominated transport. We note a correlation between this transition and the MR: the electron-dominated transport at low temperatures is accompanied by a striking positive MR (Fig. 2(b)). Since the hole contribution to transport increases with temperature, this positive MR decreases, eventually changing sign.

The transition from region I to II is signaled by this change from positive MR to negative MR. In the vicinity of the crossover temperature, the MR has a very weak field dependence (Figs. 2(b) and 2(d)), while at higher temperatures, the MR is negative and relatively small. In addition, the magnetic field dependence of $\rho _{xy}$ in region II conforms to the classic behavior characteristic of single carrier ferromagnetic metals (Fig, 2 (c)). The standard analysis of the anomalous Hall effect using Eqs. (1) and (2) may then be successfully applied by combining the measured magnetization, magneto-resistance and Hall data.  An example of such analysis is seen in Fig. 3(b) where we show the ordinary and anomalous components of the Hall resistivity; such fits yield an exponent $\gamma =1.8\pm 0.2$, consistent with similar observations in other ferromagnets.\cite{berger} The rapid increase in resistivity with rising temperature (see Fig. 1(a)) leads to an increasingly large contribution from the anomalous Hall effect; hence, the determination of $R_{H}$ from Eq. (1) is prone to large errors at higher temperatures. Thus,  extracting the sign and density of the carriers from the measurement of $\rho _{xy}$ becomes increasingly difficult with increasing temperature.

As the temperature is increased beyond T $\sim $250 K, we enter regime III which is characterized by increasingly complex behavior in both the MR and $\rho _{xy}$ as the sample passes through the Curie temperature (Figs. 2(e) and 2(f)). A clear (negative) peak in the high field MR is observed in all four samples near T$_{C}$, with the peak temperature varying slightly from sample to sample. As the temperature is increased through T$_{C}$, $\rho_{xy}$ becomes increasingly linear. However, the anomalous contribution still dominates $\rho_{xy}$ in the paramagnetic regime above T$_{C}$. Since the ordinary and the anomalous Hall effects both vary linearly with B, separation of the two contributions is no longer possible in this regime.

Figure 4 summarizes our analysis of the Hall effect in regimes I and II, showing the ordinary Hall coefficient at high fields (B $>$ 6 T); the error bars in the high temperature data reflect the difficulties mentioned earlier in separating the ordinary and anomalous contributions. The figure indicates that -- even within the uncertainty indicated -- the data show clear evidence for a crossover from electron- to hole- dominated transport with increasing temperature. This further implies the presence of two distinct carrier bands of electron- and hole-character near the Fermi surface, which is consistent with calculations of the band structure.\cite{sanvito} The presence of both electrons and holes in MnAs epilayers has direct implications for spin injection experiments since the spin scattering time for holes and electrons differs substantially in semiconductors.\cite{crookerprl96} The results presented here indicate that spin injection from MnAs into a semiconductor could involve either electrons or holes, depending upon the temperature. The dual carrier nature of this material hence offers interesting opportunities for examining both electron and hole dependent spin transport properties in the same structure.

Finally, we comment on the MR observed in the various regimes. The negative MR in regimes II and III (Figs. 2(d) and 2(f)) is qualitatively consistent with a reduction in conventional spin-disorder scattering as ferromagnetic domains are aligned by a magnetic field. The positive MR in regime I is suggestive of the contributions of two bands (electrons and holes) to the electrical transport.\cite{ssp} Similar behavior has recently been reported in the half-metallic ferromagnet CrO$_{2}$, where the crossover from hole-dominated to electron-dominated transport is also accompanied by a change in the MR from negative to positive.\cite{wattsprb} Our data suggest that temperature-induced changes in the dominant carrier species (and the accompanying differences in scattering times) are the underlying cause for the observed temperature-variations in MR. A detailed analysis of the MR and the Hall data to extract electron/hole concentration and mobility is beyond the scope of this paper and will be reported elsewhere.\cite{berryprb}

In summary, comprehensive measurements of the MR and the Hall effect in MBE-grown MnAs epilayers reveal a rich variety of electrical transport characteristics that appear to arise from the presence of both electrons and holes. The contributions of these carriers to the electrical transport vary with the temperature and applied magnetic field. A clear manifestation in Hall measurements of electron-dominated transport is observable in higher quality samples that have a RRR $>$ 50. We also observe a crossover from positive to negative MR at temperatures where an electron- to hole-dominated transition is manifest in the transport. The data presented here provide an important framework for experiments that utilize MnAs as a spin injector into semiconductors, and may also be relevant to the understanding of magneto-transport in newly developed ``digital ferromagnetic heterostructures'' wherein localized regions of MnAs are inserted into a GaAs matrix.\cite{kawakamiapl}

We thank Stefano Sanvito and Nicola Hill for useful discussions. J.J.B., K.C.K., N.S., and S.H.C. were supported by grant nos. DARPA/ONR N00014-99-1-1093, ONR N00014-99-1-0071 and -0716. P.S. and S.P. were supported by grant nos. DARPA N00014-00-1-0951, NSF DMR 97-01548 and 01-01318, and ARO DAAD 19-00-1-0125. J.J.B. also acknowledges fellowship support from the National Science Foundation.

\begin{center}
\bigskip \newpage

{\bf Figure Captions}
\end{center}

\bigskip

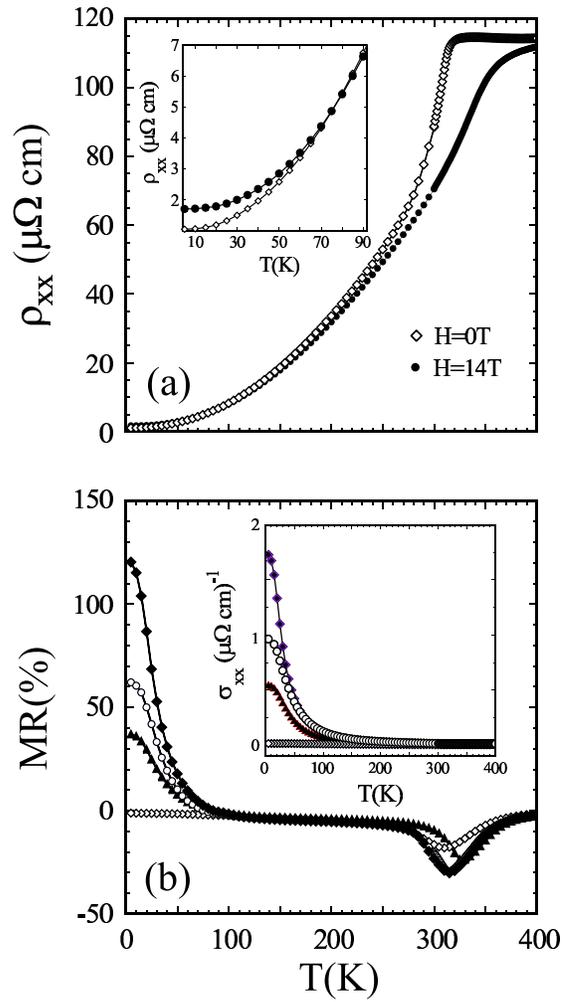
Fig. 1. a) The temperature dependence of sample resistivity for sample B at H = 0 T (open diamonds) and H=14 T (solid circles), showing a negative MR at high temperatures. Inset shows the low temperature data in more detail to highlight the positive MR at low temperatures. b) MR for samples of different RRR values; RRR = 195 solid diamonds, RRR=109 open circles, RRR=75 solid squares, RRR=5.7 open diamonds. Inset shows the zero field conductivity vs. T for these samples.

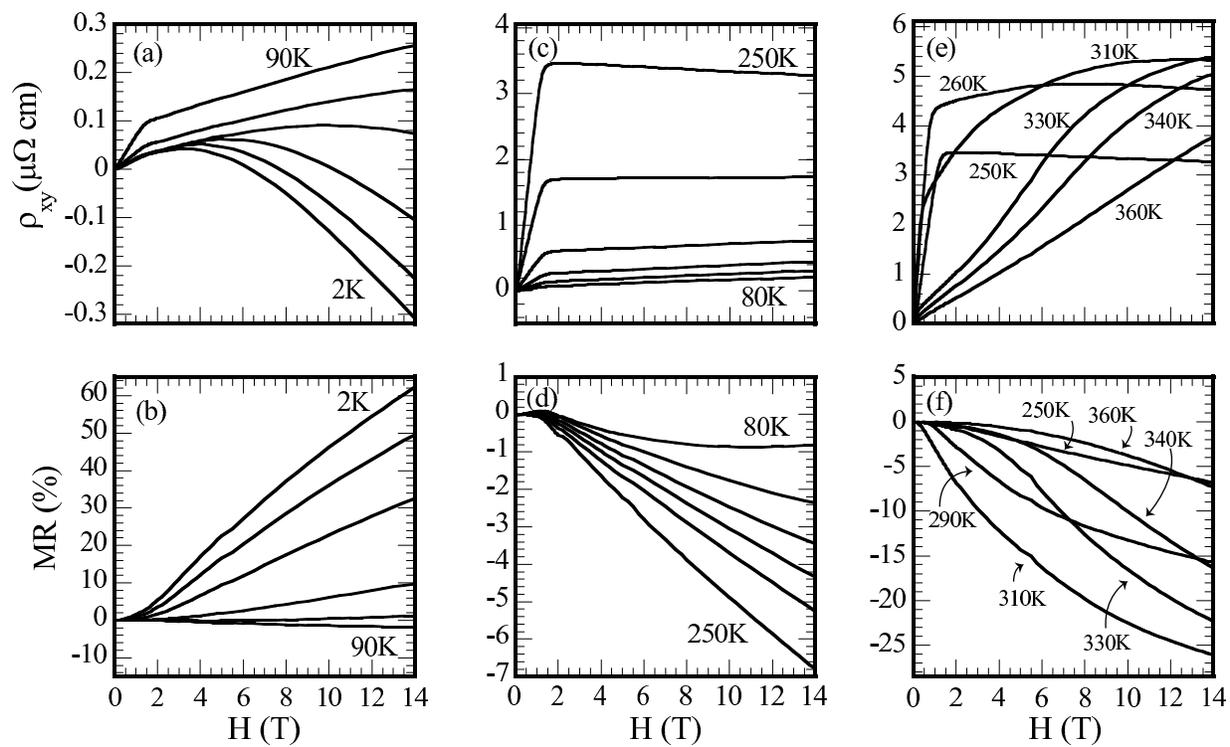
Fig. 2. The magnetic field dependence of the magnetoresistance and Hall resistivity for sample B. Panels (a) and (b) correspond to T = 2K, 20K, 30K, 50K, 70K, 90K; panels (c) and (d) correspond to T=80K, 100K, 120K, 150K, 200K, 250K; panels (e) and (f) correspond to T = 250K, 290K, 310K, 330K, 340K, 360K.

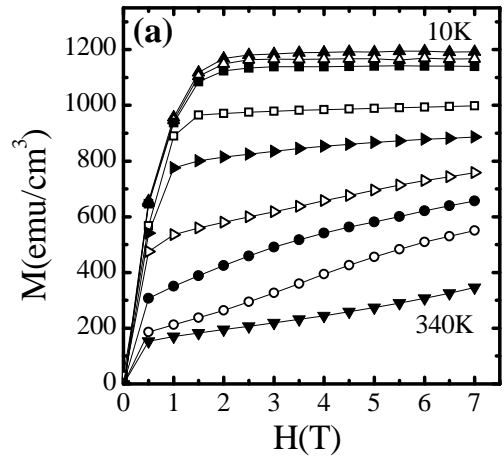
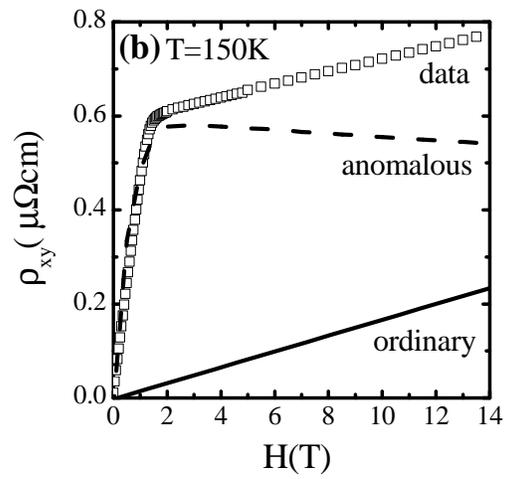
Fig.3. (a) Magnetic field dependence of the magnetization of sample B at different temperatures (T = 10K, 70K, 150K, 250K, 280K, 300K, 310K, 320K, 340K from top to bottom). (b) Decomposition of the Hall resistivity in sample B (at T $=$ 150 K) into its ordinary and anomalous components (solid line and dashed line, respectively).\cite{analysis}  The raw Hall resistivity is shown as open squares. 

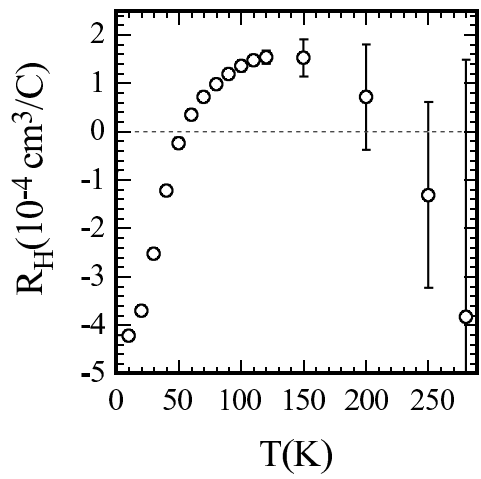
Fig. 4. The temperature dependence of the Hall coefficient in sample B, showing a change in sign of the dominant carriers near 80 K. The large error bars at the highest temperatures preclude the unambiguous identification of the dominant carrier sign for T $>$ 150K.


\begin{references}
\bibitem{analysis} Temperature independent values of the parameters $a$ and $\gamma$ (Eq. 2) are determined using the magnetization and transport data over the range T = 80 K to 250 K and H $>$ 4 T to yield linear ordinary components of $\rho _{xy}$ at each temperature.
\end{references}
\end{document}